\documentstyle[12pt,epsf]{article}
\def\beq{\begin{equation}}
\def\eeq{\end{equation}}
\def\bea{\begin{eqnarray}}
\def\eea{\end{eqnarray}}
\def\bef{\begin{figure}}
\def\enf{\end{figure}}
\def\S{{\bf S}}

\def\C{{\bf C}}
\def\Z{{\bf Z}}
\def\R{{\bf R}}
\def\P{{\bf P}}

\def\CB{{\cal B}}
\def\CC{{\cal C}}
\def\CE{{\cal E}}
\def\CF{{\cal F}}
\def\CK{{\cal K}}
\def\CL{{\cal L}}

\def\CI{{\cal I}}
\def\CN{{\cal N}}
\def\CO{{\cal O}}

\def\ba{\begin{array}}
\def\ea{\end{array}}
\def\bce{\begin{center}}
\def\ece{\end{center}}

\def\del{\partial}

\def\IC{{\relax\hbox{$\inbar\kern-.3em{\rm C}$}}}
\def\ID{\relax{\rm I\kern-.18em D}}
\def\IE{\relax{\rm I\kern-.18em E}}
\def\IF{\relax{\rm I\kern-.18em F}}
\def\IG{\relax\hbox{$\inbar\kern-.3em{\rm G}$}}
\def\IGa{\relax\hbox{${\rm I}\kern-.18em\Gamma$}}
\def\IH{\relax{\rm I\kern-.18em H}}
\def\II{\relax{\rm I\kern-.18em I}}
\def\IK{\relax{\rm I\kern-.18em K}}

\def\IQ{\relax\hbox{$\inbar\kern-.3em{\rm Q}$}}

\begin{document}
\begin{titlepage}
\rightline{SU-ITP-01/39}\rightline{HUTP-01/A044} \rightline{HU-EP-01/35}
\rightline{hep-th/0110050}
\def\today{\ifcase\month\or
January\or February\or March\or April\or May\or June\or July\or
August\or
September\or October\or November\or
December\fi, \number\year} \vskip 1cm \centerline{\Large \bf Geometric
Transition versus Cascading Solution} \vskip
1mm
\vskip 1cm \centerline{\sc Keshav
Dasgupta $^{a,}$\footnote{keshav@itp.stanford.edu},
Kyungho
Oh$^{b,}$\footnote{On leave from Dept. of Mathematics, Computer Science and
Physics, \\ \phantom{123} University of
Missouri-St. Louis, oh@hamilton.harvard.edu}, Jaemo
park$^{c,}$\footnote
{jaemo@physics.postech.ac.kr}
and Radu Tatar$^{d,}$\footnote{tatar@physik.hu-berlin.de}} \vskip 1cm
\centerline{{ \it $^a$ Department of Physics,
Stanford University, Stanford CA 94305-4060, USA}} \vskip 1mm
\centerline{{ \it $^b$ Lyman
Laboratory of Physics, Harvard University, Cambridge, MA 02138, USA}}
\vskip 1mm
\centerline{{\it $^c$ Department of Physics,
POSTECH, Pohang 790-784, Korea}}
\vskip 1mm
\centerline{{\it $^d$ Institut
fur Physik, Humboldt University, Berlin, 10115, Germany}} \vskip 2cm
\centerline{\sc Abstract} \vskip 0.2in

We study Vafa's geometric transition and Klebanov - Strassler solution
from various points of view in M-theory. In terms of brane
configurations,
we show the detailed equivalences between the two models. In some limits,
both models have an alternative realization as  fourfolds in M-theory with
 appropriate G-fluxes turned on. We discuss some aspects of the
fourfolds including how to see the transition and a possible extension
to the non-supersymmetric case.

\vskip 1in \leftline{October 2001}
\end{titlepage}
\newpage
\section{Introduction and Summary}
The large $N$ limit of ${\cal N} = 4$ conformally
invariant theory has a supergravity
dual which is used to study many aspects of this theory
\cite{malpolwit}. This can be also be extended to theories
with lower supersymmetry. The supergravity duals are now on more exotic
backgrounds like orbifolds and conifolds. However the situation is more
subtle for non-conformal theories with lower supersymmetry. When we have
$N$ D3 branes near a conifold singularity, the world volume theory is
${\cal N}=1$ $SU(N)\times SU(N)$ theory with a quartic superpotential. To
break conformal invariance in these theory we have to introduce fractional
branes \cite{kn,ot1}.
Fractional branes are higher dimensional branes wrapped on vanishing
cycles of a manifold and therefore they carry charges of lower dimensional
branes. At the conifold points there are vanishing two cycles on which
D5-branes of type IIB theory can wrap. The fractional charges of these
branes are generated by $\int B_{NSNS}$ fields that thread through the
vanishing cycles. In the presence of $M$ such fractional branes and $N$
integer branes (which are of course D3 branes) the world volume theory is
a non-conformal ${\cal N} =1$ $SU(N+M)\times SU(N)$ theory in the UV.
This is basically
the {\it Klebanov-Strassler} model\cite{ks}.

There is yet another way to generate a non-conformal ${\cal N}=1$ theory
in four dimensions. This is by wrapping a D5 brane on the finite two-cycle
of a resolved conifold.
On the world volume of the D5 brane there is an IR theory (on the remaining
unwrapped $3+1$ dimensions) with a superpotential which can be calculated
using geometric engineering. The superpotential breaks the ${\cal N} = 2$
theory to ${\cal N}=1$. This is {\it Vafa's} model\cite{vafa} (see also
\cite{civ,eot,ckv}). In the UV therefore it is six dimensional whereas the
Klebanov-Strassler model remains four dimensional at UV.

Both these theories have a {\it dual} picture at all scales
where we have a deformed
conifold with no branes. The branes are replaced by three form fluxes. The
three form RR fluxes are of course the remnant of the wrapped D5 branes. The
NS fluxes in Klebanov-Strassler have their origin from the $B_{NSNS}$ fluxes
through the vanishing two cycle which is related to the gauge coupling of the field theory. For Vafa's model the origin of NS fluxes is a
little subtle. It is related to the ${\it size}$ of
the resolved $S^2$ which gives the gauge coupling of the field theory
\footnote{Other reasons for the existence of $H_{NS}$ fluxes in Vafa's model
have been discussed in \cite{dot,dot1}.}.
Since we don't expect the NS fluxes to be constant, therefore the size of
the two cycle is also not a constant quantity. In fact as was shown in
\cite{vafa,dot,dot1} the size of the two cycle determines the RG flow of the
$3+1$ dimensional gauge coupling.

In Klebanov-Strassler model the theory flows in the far IR to a system with
only fractional branes
via consecutive {\it cascades} which in field theory are Seiberg dualities.
As we discussed above there is, at all scales, a dual
non-singular closed string background with only the three form fluxes.
In Vafa's model the corresponding dual can be reached by a
{\it conifold transition}.
Both of these models have yet another M-theory or IIA dual
via a T-duality in which we get a picture completely in terms of
brane constructions\cite{dot,dot1}. In this way its very easy to see why
we go to a closed string background. A detailed analysis of this will be
presented in sec. 3.3.

The brane realization of the closed string background as given in
\cite{dot,dot1} can be used to our advantage to actually derive the
supergravity solution. We will see how far we can trust the supergravity
solution obtained by this method. This will be shown in sec. 4.2 as example 2.
We shall point out that due to {\it delocalisation} this in fact doesn't
give us the exact answer which can nevertheless be derived via a different
technique.

The technique that we shall use has been discussed earlier in a different
context in \cite{bbsisters,gukov,drs,guk} and in a recent related context in
\cite{gubser,granopol1,granopol2,kachru} (see also \cite{maycuriomayr} for
related issues). Herein we take a fourfold
(which we shall assume to be compact for our purpose)
and switch on an appropriate G-flux. The fourfold
that we take is a non-trivial torus fibration over a base which is a
deformed conifold. The G-fluxes have one leg along the fiber and the other
legs on the base $-$ deformed conifold. This background in M-theory is related
to the closed string background of the above two models in IIB. There are
however some subtlety related to the global structure which we will point
out in sec. 4.3. We first give a warm up example in sec. 4.2 (as example 1)
which will illustrate the basic procedure. In the next section we give
the solutions to the model. We shall see that from M-theory we directly get
the {\it linear} forms of the background supergravity equations of motion. The
solutions of these equations will predict the behaviour of these models at
all scales.

The study of
the Klebanov-Strassler's and Vafa's models from fourfold points of view
has an inherent advantage. The background can be easily extended to the
non-supersymmetric case. Such non-supersymmetric background was recently
studied in \cite{bbsis,haack}.
However its not yet clear how the dual brane side would look like.
We discuss all these issue in sec. 5. We conclude with some comments on other
issues like a possible extension to the Type I background and fate of the open
strings in sec. 6.

We now begin with a discussion of Vafa and Klebanov-Strassler models.

\section{Vafa's Geometric Transition for $U(N)$ theory}
\subsection{General Features for the Transition}
In \cite{vafa}, based on Chern-Simons/topological strings duality
\cite{gova},
a duality transition was proposed between $N$ D5 branes wrapped on the
finite
two cycle of a resolved conifold and a geometrical picture consisting
of a
deformed conifold  without branes but with a $H_{RR}$ through
the
finite three-cycle of the deformed conifold and  $H_{NS}$ flux through
the
noncompact three-cycle. As explained in \cite{vafa} and further clarified
in
\cite{civ}, the exact match between the parameters of the field theory
on the
$N$ D5 branes and the parameters read from the geometry is:

$\bullet$ The flux of $H_{RR}$ through the $\S^3$ cycle is equal to
$N$, the
number of the D5 branes which disappear in the geometry.

$\bullet$ The flux of  $H_{NS}$ through the noncompact 3-cycle of the
deformed
conifold is equal to $\alpha$, the bare coupling constant of the field
theory
living on the D5 branes.

$\bullet$ The period $S$ of holomorphic 3-form on the  $\S^3$ cycle is
equal to
the gluino condensate on the field theory.

$\bullet$ The quantum corrections (RG flow) of the field theory on the
D5
branes are related to the the period $\Pi$ of holomorphic 3-form on the
noncompact 3-cycle in the deformed conifold side. Because this period
is over
a non-compact cycle, \cite{civ} have imposed a cutoff in the integral
for
large distances in the Calabi-Yau manifold (IR cutoff in the
Calabi-Yau).
This cutoff was the identified with a dynamical scale of the $U(N)$
theory
which is an UV cutoff. To do so, one has to remember the usual UV/IR
correspondence for branes.

$\bullet$ The superpotential in the $\CN = 1$ field theory is
determined
by the fluxes in the geometry:
\bea
-\frac{1}{2~\pi~i} W_{eff}~=~N~\Pi~+~\alpha~S
\eea
and after the previous identifications becomes
\bea
W_{eff}~=~-N~S~\mbox{log}~S+~S~(3~N~\mbox{log}~\Lambda_0~-~2
\pi~i~\alpha)
\label{supot1}
\eea
Furthermore, after identifying the cutoff $\Lambda_0$ with the UV
cutoff in
the field theory, \cite{vafa,civ} obtained the usual superpotential for
the
glueballs
\bea
W_{eff}~=~ N~S{\Big (}log (\Lambda^3/S) + 1{\Big )}
\label{supot2}
\eea
\subsection{Vafa's Transition in MQCD}
In \cite{dot,dot1}, the transition of \cite{vafa} was discussed in
terms of
transitions in MQCD. By starting with $N$ D5 branes wrapped on the
$\P^1$
cycle of the resolved conifold, the T-dual picture will be a brane
configuration with $N$ D4 branes along the interval with two NS branes
in the `orthogonal' direction at the ends of the the interval.
Here the length of the interval is the same as the size of the
rigid $\P^1$.  As the rigid $\P^1$ shrinks to zero, the size of
the interval goes to zero and the two NS branes approach each other.
To ease the next discussion, let us denote the common 4 directions for
all the
branes as $(x^0,~x^1,~x^2,~x^3)$, the extra direction of the D4 brane
by
$x^n$ ($n$ comes from noncompact, it will be clear below why we
use this
notation) and the extra directions of the two orthogonal NS branes are
$(x^4,~x^5)$ and $(x^8,~x^9)$ respectively.

When we lift this configuration to M-theory,
the two NS5 branes become M5 branes and
are connected together by another M5 brane emanating from the D4
branes. Defining complex coordinates as: \bea\label{complexcoord}
x = x^4+ix^5, ~~~y=x^8+ix^9,~~~ t = exp(-R^{-1}x^{n}+ix^{10})\eea
where $R$ is the radius of the $11th$ direction, the world volume
of the M5 corresponding to the resolved conifold is given by
$R^{1,3} \times \Sigma$ and $\Sigma$ is a complex curve defined,
 up to an undetermined constant $\zeta$, by
\bea \label{m5res} y = \zeta x^{-1}, ~~t = x^{N}\eea

As explained in \cite{dot,dot1}, when the size of $\P^1$ goes to
zero, the $x^n$ direction becomes very small
and the value of $t$ on $\Sigma$ must be constant because
 $\Sigma$ is holomorphic and there is no non-constant holomorphic map
into $\S^1$. Therefore the  M5 curve makes a transition
 from a ``space'' curve into a ``plane'' curve. From (\ref{m5res}), we
obtain two relation on $t$ and $t^{-1}$
 \bea \label{rel} t= x^N,~~ t^{-1} = \zeta^{-N} y^N.
 \eea
So there are $N$ possible plane curves which the M5 space curve
$\Sigma$ can be reduced to: \bea \label{m5plane}
\Sigma_k:~~~t=t_0, ~~xy = \zeta \exp {2\pi i k/N}, \quad k =0, 1,
\ldots , N-1. \eea

This is thus the way we see Vafa's duality transformation. After
the transition the degenerate M5 branes are no longer considered
as the M-theory lift of D4 branes. This is now a closed string
background.

\section{Klebanov-Strassler Cascading Solution}
\subsection{Cascading Solution}
In \cite{ks}, an interesting approach was taken to study the conifold
with integer $N$ D3 branes and $M$ fractional D3 branes. Using results
of
\cite{kn}, the gauge group in the UV is $SU(M+N) \times SU(N)$ and the
field
theory flows to infrared by an RG flow which will give a scale
dependent
number of colors
\bea
N(r)~=~N_0~+~\frac{3}{2~\pi}~g_s~M^2~\mbox{ln}(r/r_0).
\eea
where $g_s$ is the string coupling constant and $r$ is the radial
direction
for the conifold. Along the RG flow the gauge group is
$SU(M + N(r)) \times SU(N(r))$ and as $\mbox{ln}(r/r_0)$ decreases by
$\frac{2~\pi}{3~g_s~M}$, $N(r)$ decreases by $M$, so we will have:
\bea
SU(M+N) \times SU(N) \rightarrow SU(N) \times SU(N-M)
\eea
and so on until the gauge group becomes $SU(M)$.
In terms of branes, this means that by starting with  $N$ D3 branes and
$M$
fractional D3 branes, along the RG flow the number of integer branes
decreases
and in the end we remain with only the $M$ fractional D3 branes. This
is the
cascade of Seiberg dualities.

In the supergravity, the $N$ integer D3 branes are the sources for the
RR
5-form which is
\bea
\tilde{F}_5 = \cal{F} + * \cal{F}
\eea
where $\cal{F}$ is proportional to $N(r)$ multiplied by the volume of
$T^{11}$ which is the horizon for the conifold.

The $M$ fractional D3 branes are sources for a RR 3-form through the
3-cycle $\omega_3$ of $T^{11}$
\bea
F_3 = M \omega_3
\eea
and the NS field through the 2-cycle $\omega_2$ of  $T^{11}$
\bea
B_2 = 3g_s M ~ \mbox{ln}(r/r_0) \omega_2
\eea

\subsection{The Transition in the Klebanov-Strassler Case}
We now describe the transition corresponding to the
Klebanov-Strassler case. We start with the conifold geometry and we
consider a
vanishing 2-cycle at the apex where we wrap $M$ D5 branes which are the
fractional D3 branes. In this case, we need to have a NS field through
the
vanishing $\P^1$ cycle in order to insure its stability.

To discuss the transition, we use the approach of \cite{dot,dot1}
modified to
be fitted for the conifold instead of the resolved conifold. We take a
T-duality in the direction $\psi$ of the conifold which we denote by
$x_c$ in
order to signal the fact that its compact. The integer D3 branes would
become
D4 branes with one direction on the circle $x_c$. As discussed in
\cite{dm1},
the fractional D3 branes become D4 branes on a half-circle. So in this
case
the D4 branes will have the 4-th spatial direction in a compact
direction.
The field theory on the $M$ fractional D3 branes is
pure ${\cal{N}} = 1,~ U(M)$
super Yang Mills theory.

After lifting the brane configuration to M theory and studying the
transition
as in \cite{dot,dot1}, we arrive to a geometrical configuration which
should
be identical to the Klebanov-Strassler case discussed in the previous
subsection. One important observation is the following: because we
started
with a conifold with a vanishing but stable 2-cycle, we have the
compact
direction $x_c$ which remains compact on the geometrical side.

In order to discuss this in more detail we use the results of
\cite{dm,dm1}
where the elliptic model was lifted to M theory.
We now define again the complex coordinates $x,~~y$ as in the resolved
conifold case and $t = exp(-R^{-1}x^{c}+ix^{10})$
where $R$ is the radius of the $11th$ direction and $x_c$ is the
compact
direction, $t$ is not periodic in $x_c$ so we should use it only for a
finite
range of values of $x_c$. The  form of the single M5-brane is
\bea
y = \zeta x^{-1}, ~~t = x^{M}.
\eea

However before we go in more details we should ask what the distance
between the branes represent. In the usual case the coupling constant
of $\CN =1$,  $3+1 D$ gauge theory is determined by the distance between
the branes. In the wrapped brane picture the coupling constant comes
from the
$B_{NSNS}$ field on the vanishing two cycle of the conifold from the
coupling
\bea
\int_{\Sigma}B_{NSNS}~\int F\wedge *F + \int_{\Sigma}B_{RR}~\int F
\wedge F
\eea
where $\Sigma$ is the vanishing two-cycle. For our case we put
\bea
B_{RR}=0,~~~~~B_{NSNS} = 3g_sM ~ ln(r/r_0) \omega_2.
\eea
The above formula implies that now the distance between the branes
is actually a function of $r$. Thus the two NS5 branes are
now bent along $r$. In the M-theory lift, the curved M5 brane model
will now have $x,y$ as functions of $r$.

If we now shrink the distance between the two M5 branes, from the
discussion of \cite{dot,dot1} we expect $t$ to be fixed to a constant
value.
In terms of type IIA picture there are now two important
considerations.

$\bullet$ The two M5 branes are now at a point $t_0$ and satisfying the
equation $xy=\zeta$. When we reduce this to IIA the bending of the M5
branes along $x^{10}$ direction will appear as 2-form field. Making
a T-duality along $\psi = x_c$ will give us a deformed conifold with
$H_{RR}$ on the 3-cycle. Observe that after
reducing to the type IIA theory we obtain an {\it single}
NS5 brane which is wrapped
on the $S^2_1 - S^2_2$ of $T^{11}$\cite{dm1}.

$\bullet$ The M5 along $x$ and $y$ directions are
 also bent along $r$. After
the transition this bending will remain in the planar M5 model.
Now under a T-duality the bending will appear as $H_{NS}$ field on a
3-cycle dual to the 3-cycle on which there is $H_{RR}$.
We thus see how
 the RR and NS forms appear in the brane picture. This is also
discussed
in some detail from supergravity perspective in
 eqt. (3.5) of \cite{mt} and
transparency 18 of \cite{klebanov}.

Therefore we see that after the transition we reach a non-singular
manifold
which is a deformed conifold. The argument is the same as discussed in
\cite{dot,dot1}.

There are two further indications that this should be the case:

$\bullet$ In \cite{dot,dot1} we have also considered the reverse
transition,
from the geometry to the brane configuration and the D4 branes appeared
due to
the Hanany-Witten effect in the presence of a singularity at the
intersection
of orthogonal NS branes. We have discussed the transition from the
deformed
conifold to resolved conifold, where the 4-th spatial direction of the
D4
branes is an interval in a noncompact direction. However there is an
ambiguity here. At the conifold point $-$ wherein we have the creation of
a D4 brane, $-$ there are now two distinct possibilities to stretch the D4
brane.
We could either stretch it along $x_n \equiv x^7$ or along $x_c \equiv
x^6$.
These two possibilities give rise to the Vafa and Klebanov-Strassler
models respectively. In the former case we have D5
branes wrapped on the $S^2$ of a resolved conifold. And in the latter
case we
have D5 wrapped on vanising $S^2$ with $B_{NSNS}$ flux.

$\bullet$ In section 5.3 of \cite{ks}, the discussion concerned the
validity
of a description for the pure ${\cal{N}} = 1, U(M)$ super Yang Mills
theory.
In order to have a reliable dual of the pure glue theory, one needs to
take
the limit of finite $g_s~M$ which is the limit when the $\S^3$ at the
apex
is finite which is exactly what we discussed above
$-$ a finite $\S^3$  with non zero
$H_{RR}$ through it. Considering the fact that we have started with
$U(M)$ theory on the fractional D3 branes, we expect to obtain
pure ${\cal{N}} = 1, U(M)$ super Yang Mills theory and this is in
accordance
to the claims of \cite{ks}.

When we reduce them to type IIA, they will be D4 branes wrapping the
$x_c$
cycle. But this is just the result of \cite{mt,klebanov} after taking
direct
T-dual to the \cite{kn} model! The corresponding D4 branes will
contribute to
the 4-form which is the T-dual to the 5-form in type IIB.

\subsection{Brane Realisation of Cascade vs. Geometric Transition}

We can also clarify now the differences between the two
approaches of studying $\CN=1$ duality. In one case $-$ of
Klebanov-Strassler
$-$ we have a T-dual model described on a {\it torus} parametrized by
$z = x^6 + i x^{10}$. The two M5 branes are at two arbitrary points on
the
$z$ torus. The transition from ``curved'' M5 to ``plane'' M5 can be
achieved
by a spiral motion on $z$ plane. This spiral motion is the brane
realization of cascade in this model. In the other case $-$ of Vafa $-$
the
T-dual model is defined on a {\it cylinder} parametrised by $w = x^{7}
+
i x^{10}$. There is no spiral motion because this description is in the far
IR\footnote{This however doesn't mean that there is no
cascade in this model. We shall discuss this in detail in a forthcoming
paper. See also \cite{vafatoday} where it is discussed in detail how
cascades occur in the geometric transition model. In the presence of
$N$ D3 branes and $M$ wrapped D5 branes the theory is $SU(N+M)\times SU(N)$.
The cascades in this theory are realised by an {\it infinite} sequence of
flop transitions. At the end $-$ which is infact far IR $-$ the theory
undergoes a geometric transition via the usual conifold transition. The brane
construction that we discuss in this paper only describes the IR aspect of
Vafa's geometric transition picture. In the presence of D3 branes we have to
consider a pair of $D5 - {\bar {D5}}$ with {\it fluxes}
wrapped on the resolved two cycle
of the conifold. The T-dual picture would now be on a compact circle. The
behaviour of tachyon in this system and other details would be discussed in
the forthcoming paper. We thank the referee for his comments.}.
Therefore, as we discussed in detail in \cite{dot,dot1},
to see the
RG flow of $\CN=1$ coupling we make a transition to the ``plane'' curve and
impose a {\it IR cutoff} on the
integral
\bea \int_{\Lambda_0^{3/2}>|x|>|\zeta|^{1/2}}~dlogx \wedge *dlogx \eea
This takes the form of
the
NSNS flux on the deformed conifold side under a T-duality.

We could be a little bit more precise here. From the above paragraph
one
would naively assume that the difference between Vafa and
Klebanov-Strassler
is because one of the model is on a torus and the other is on a
cylinder.
However this is {\it not} the case. To see this let us first consider
Vafa's geometric transition:

The M-theory curve is
\bea
t= x^N, ~~~ t^{-1}=\zeta^{-N}y^N
\eea
At the transition point $-$ when the two M5 brane ($x$ and $y$) come at
a
point $x^7=0$ $-$ there is a $S^1$ from the M5 between them. As
discussed in detail in \cite{dot,dot1} this is not holomorphic and
therefore the only holomorphic curve is when $x,y$ satisfy
$xy=\zeta$ at a point $(x^7=0, t=t_0)$.

Now we shall argue that something more interesting happens
for Klebanov -
Strassler
model because we also have D3 branes.
To delve into it we would need some details of M-theory curves.
When we have $N$ D3 branes at the apex of a conifold then the M-theory
lift (of the T-dual) is given by a curve
\bea
x^N = y^N = 0
\eea
This means that the D4 branes which turn into M5 branes go right
through the
other two (orthogonal M5). This M5 we call as toroidal
M5\footnote{Because of
this construction the $SU(N)\times SU(N)$ theory has a global symmetry
of
$SU(2) \times SU(2)$ from the rotation of the two ``decoupled'' M5
branes.}.

Also when we have D5 wrapped on {\it vanishing} $S^2$ the M-theory
lift of it is a MQCD like structure. However when we have {\it both}
i.e. $N$ D3 and $M$ D5 wrapped on $S^2$ then the M-theory lift would be
a toroidal M5 and a MQCD five brane.

{}From this construction its now easy to see what happens when we
have $x^6=0$. The MQCD five brane comes to a point. But now there is no
$S^1$ there. Because of the existence of toroidal M5 branes the system
remains
holomorphic even when we make $x^6=0$. Therefore in the original model
$-$ when we have $N$ D3 branes $-$ the UV description of the metric in
\cite{kn} is the same as that of \cite{ks}.
This is consistent with the predictions of
Klebanov- Strassler. Question is what happens at the infrared.
Motivated by the analysis done above for the case of Vafa, we expect
a similar ``transition''
when the toroidal M5 can be made to
go away. This would happen when we make one of the M5 $x$ move spirally
up
on the $x^6 + i x^{10}$ torus. Everytime the M5 cross the other one
there
is a Seiberg duality in the theory and the number of D3 reduces. This
is the cascade. Ultimately when the D3 branes completely go away then in
M-theory
making $x^6=0$ will lead to a holomorphic structure when we also make
$t=t_0$. This is now the transition.

\section{Geometric Transitions from G-Flux}

\subsection{Basic Idea}

In the previous sections we have seen how Klebanov-Strassler and Vafa's
transitions are realized from brane constructions in M-theory. The key
point in the constructions was the existence of holomorphic structures.
Demanding holomorphicity {\it after} the conifold transition gives us
essentially the closed string background with fluxes and no branes.

There is yet another realization of the transition from M-theory and
this
is by invoking the idea of G-fluxes. The whole process can be
summarized
by the following steps:

$\bullet$ Consider a 4-fold in M-theory which is a non-trivial $T^2$
fiberation over a base $\cal B$.

$\bullet$ Switch on a G-flux which has one component along the $T^2$.
Reducing to type IIB, the G-flux will give the NSNS and RR three form
fields on the base $\cal B$.

$\bullet$ If we choose $\cal B$ to be a deformed conifold then this
will
effectively produce one side of the geometric transition, i.e the side
with a closed string background with fluxes and no branes\footnote{The global
structure, as we will discuss later, is more complicated when the fiberation
is non-trivial. In this section and the next we will not dwell into this
subtlety even though we take non-zero Euler-characteristics, $\chi$, of the
fourfold.}.

$\bullet$ The presence of G-fluxes the 4-fold metric will be warped and the
warp factor depends essentially upon the Euler characteristics of
the 4-fold and is related to a hierarchy of energies in the dual field theory
\cite{kachru}.

$\bullet$ By doing a conifold transition on the base, we should be able to
argue that the warped metric now gives the metric of a D5 wrapped on
a two cycle of a resolved conifold\cite{malnun}.
This would then signify Vafa's
transition at least in the far IR.

For the case of Klebanov-Strassler the M-theory realization is been
studied in some details in \cite{gubser,kachru}. For example in
\cite{gubser} it was shown that a fourfold which is a direct product
of a deformed conifold and a torus actually do not realize the required
background as the Euler characteristics is zero and, in the absence of
any D3 branes, the quantity $\int G\wedge G$ also vanishes.

In \cite{kachru} the Klebanov-Strassler model has been embedded in a
F-theory
compactification. The Calabi-Yau fourfold $X$ which admits a conifold
singularity in its base $\cal B$ is given by specifying a Weierstrass
model
\bea
y^2 = x^3 + x f(z_i) + g(z_i)
\eea
where the base $\cal B$ is given by
a quartic equation in $P^4$
as:
\bea
P \equiv z_5^2(\sum_{i=1}^4z_i^2) - t^2z_5^4 + \sum_{i=1}^4 z_i^4 = 0
\eea
In the above equation $z_i$ are the homogeneous coordinates on $P^4$
and $t$
a real parameter. $f$ and $g$ are polynomials of degree 4 and 6 in the
homogeneous coordinates $z_i$. The fourfold is realised here as a
non-trivial
$T^2$ fiberation over a conifold base. As such its Euler characteristics
can
be shown to be ${\chi\over 24} = 72$ \cite{kachru}.

To realize the geometric transition of Vafa we need a fourfold which is
a
non trivial $T^2$ fibration over a deformed conifold base. In the next
section
we will give an example of a fourfold which is a non trivial $T^2$
fiberation
over a base $T^6/ Z_2 \times Z_2'$ where the $Z_2$'s are orientifold
actions.
To start with, let us see whether we can say something
about the
case when we have a fourfold with base a conifold and the fluxes
switched on.
Determining the exact background is a difficult exercise but we can use
various approximate methods to get a possible solution. The conifold
base
(in the absence of any fluxes) is given by the familiar equation\footnote{This
is the conformal transformed metric where ${dr\over r} \equiv d\phi$. Using
this form of the metric the base of the conifold can be easily shown to be
Einstein\cite{candelossa}.}
\bea
{dr^2\over r^2} + \sum_{i=1}^2~(d\theta_i^2+sin^2\theta_id\phi_i^2)
+ (d\psi + cos\theta_1d\phi_1 + cos\theta_2d\phi_2)^2
\eea
For the case of $\theta_2 = \phi_2=$ constant, the above equation is
just
an ALE space and therefore should be given by a D6 brane when the
M-theory
radius is very small. In fact as argued in \cite{kochoh} two
intersecting
D6 branes in IIA when lifted to M-theory is actually a conifold\footnote{There
is an interesting digression. Two intersecting D6, with four common directions,
 at an angle (in the presence
of O6 planes) realize a {\it seven} dimensional $G_2$ holonomy manifold
when the system is lifted to M-theory. Using this construction and applying
the methods of \cite{blu1,blu2,blu3}, one can study
chiral matters in ${\cal N} =1$ theory in four-dimensions
\cite{garyshu,garytwo,atiwit,witg2,bobby}. We shall however not discuss this
interesting connection in this
paper.}.
However the
above discussions are in the absence of any fluxes. In the presence of
fluxes the background metric is complicated but can be worked out.

If $r$ is the radial direction of the conifold base, $x^1$  the
M-theory
direction and the torus $T^2$ parametrised by $x^1,x^2$ then for small
$r$ the torus has a very small warp factor given by
\bea
{c_1^2+1\over c_1^2+H} (dx^1)^2 +{c_2^2+1\over c_2^2+H} (dx^2)^2
\eea
which is effectively 1 near $r = 0$.
The factors $c_i, i = 1,2$ are the value of $C$ fields
$C_{1\theta_1\psi}, C_{2\theta_2\psi}$ at the origin $r=0$. And $H$ is
a
linear function of $r$. The quantities $\theta,\psi$ are defined before
and they are coordinates for a conifold.

Therefore, the above metric is an approximate fourfold where the
fiberation
is  trivial and there is a $H_{NSNS} + \tau H_{RR}$ over a
three cycle whose cohomology is given by
\bea
e^{\psi}\wedge e^{\theta_1} \wedge e^{\phi_1}-e^{\psi}\wedge
e^{\theta_2} \wedge e^{\phi_2}
\eea
where $e^{\theta_i}=d\theta_i, e^{\phi_i}=sin\theta_i~ d\phi_i,~
i=1,2$. The
Euler characteristics of this fourfold is zero as the fiberation is
trivial.
In the next section we shall discuss a fourfold which has a non-trivial
$T^2$ fiberation.

\subsection{A First Look: Delocalized Case}

{\it Outline of the Setup}

Consider a 4-fold given by a non-trivial $T^2$ fiberation over a base
${\cal B}$. The $T^2$ is parametrised by $x,y$ such that
\bea
dz = dx + \tau dy, ~~~~~d{\bar z} = dx + {\bar \tau}dy
\eea
where $\tau$ is the complex structure of the torus. We now choose a
G-flux
in the following way:
\bea \label{choiceg}
{G\over 2\pi} = dz \wedge \omega - d {\bar z} \wedge *\omega
\eea
where $\omega \in H^{1,2}({\cal B})$. The flux then lifts to a
combination of
the NS field strength $H_{NSNS}$ and RR field strength $H_{RR}$. This
is
given by
\bea \label{fluxname}
H_{NSNS} = \omega - *\omega, ~~~~~H_{RR}= \omega\tau - *\omega{\bar
\tau}
\eea
Before we go any further let us remind ourselves of the following
important conditions \cite{sethi,bbsisters,gukov,drs,guk}:

$\bullet$ The 4-fold vacua has a tadpole anomaly given by $\chi /24$
where
$\chi$ is the Euler characteristics of the 4-fold.  If $\chi /24$ is
integral, then the anomaly can be canceled by placing a sufficient
number
of spacetime filling branes $n$
 on points of the compactification manifold. There
is also another way of canceling the anomaly and this is through the
G-flux.
The G-flux contributes a $C$ tadpole through the Chern-Simons coupling
$\int C\wedge G \wedge G$. When  $\chi /24$ is not integral then we
need
both the branes and the G-flux to cancel the anomaly. The anomaly
cancellation
formula becomes
\bea
{\chi \over 24} = {1\over 8\pi^2}\int G \wedge G + n
\eea
which must be satisfied for type IIA or M-theory.

$\bullet$ If we denote the spacetime coordinates by $x^{\mu}$ where
$\mu
= 0,1,2$ and the internal space by the complex coordinates $y^a, a = 1,..,
4$
then in the presence of G-flux the metric becomes a warped one
\bea
ds^2 = e^{-\phi(y)}\eta_{\mu\nu}dx^{\mu}dx^{\nu}+ e^{{1\over 2}\phi(y)}
g_{a{\bar b}}dy^ady^{\bar b}
\eea
with the G-flux satisfying the condition
\bea
G = *G, ~~~~~ J \wedge G = 0
\eea
where the Hodge star acts on the internal 4-fold with metric $g$ and
$J$ is
the Kahler form of the 4-fold. There is also another non vanishing G
given
in terms of the warp factor as $G_{\mu\nu\rho a} =
\epsilon_{\mu\nu\rho}
\del_a e^{-{3\over 2}\phi}$. The warp factor satisfy the equation
\bea \label{wareq}
\Delta e^{3\phi \over 2}= *[4\pi^2 X_8 - {1\over 2} G \wedge G -
4\pi^2 \sum^n_{i =1} \delta^8(y - y_i)]
\eea
where the Laplacian and the Hodge * is defined wrt $g$, and $X_8$ is
the 8-
form constructed out of curvature tensors.

$\bullet$ The anomaly cancellation condition will now become, in type
IIB
theory
\bea
{\chi \over 24}= n + \int H_{RR} \wedge H_{NSNS}
\eea
where $n$ is the number of D3 branes. Observe that if we choose the
right
background fields satisfying $n=0$ we have the background required
for the Vafa and Klebanov-Strassler case.

{\it Various Notations and Scales}

Before we go any further let us discuss the various limits that we
need to impose on the scales to get a supergravity background for our
case.
Let us denote the eleven dimensional Planck length by $l_{11}$ and the
average volume of the fourfold to be $l_8^8$. We shall assume that
$l_8 >> l_{11}$. The background G-flux and the warp factor
$\omega = e^{\phi}$
scales as
\bea G = [l_{11}^3/l_8^4],
~~~~~ \omega = e^{[l_{11}^6/ l_8^6]}\eea
For very large sized fourfold the metric becomes unwarped and the
background G-flux vanishes.

For finite sized fourfold there is a warp factor which is determined by
eq. (\ref{wareq}). In terms of derivative expansions both the
$X_8$ and the membrane terms are suppressed by ${1\over l_{11}^6}$ and
therefore can be neglected compared to the  $G \wedge G$ term
\cite{drs}.

The other $R_{MNPQ}^4$ terms in the low energy M-theory lagrangian
are also suppressed  by ${1\over l_{11}^6}$.
These terms are written in terms of $\epsilon$ and $t_8$ tensors of
\cite{tseytu}.

Our definition for the Hodge * in $d$ dimensions will be:
\bea
*(dx^{a_1}\wedge .. \wedge dx^{a_q}) = {1\over (d-q)!}
\epsilon^{a_1...a_q}_{a_{q+1}...a_d}~dx^{a_{q+1}}\wedge ..\wedge dx^{a_d}
\eea
where $a_p$ are real coordinates. In terms of complex coordinates $a,b$
the epsilon tensor can be expressed as:
\bea \epsilon_{abcd{\bar e}{\bar f}{\bar g}{\bar h}} = g_{a{\bar e}}
g_{b{\bar f}}g_{c{\bar g}}g_{d{\bar h}}\pm permutations \eea

We will also assume that the number $M$ of wrapped branes (on either vanishing
cycles or finite cycles) to be very large such that even for the case
$g_s \to 0$ ($g_s$ being the string coupling) the quantity $g_sM$ is very
large. In the UV therefore, for the Klebanov-Strassler model, we assume
\bea \label{kslimit}
g_s \to \epsilon,~~ \int_{S^2\to 0} B_{NSNS}\to \epsilon^{-\beta},
~~ \beta > 1 \eea

{\it Example 1}

As a warm up example consider a 4-fold given by $T^8/G$ where $G$ is
the
orbifolding group. Therefore type IIB theory will be compactified on
the
orientifold $T^6/Z_2 \times Z'_2$ and $Z_2$ involves the orientifold
group.

In M-theory we define the G-flux to be:
\bea
G = Ad{\bar z}^1dz^2d{\bar z}^3dz^4+Bdz^1d{\bar z}^2dz^3d{\bar z}^4
+Cd{\bar z}^1dz^2dz^3d{\bar z}^4+Ddz^1d{\bar z}^2d{\bar z}^3dz^4
\eea
where $z^i$ are the complex coordinates of $T^8$ and $z^4$ will be the
direction along which we reduce to go to type IIB. The constants
$A,B,C,D$
are related by the identity
\bea
AB + CD = {\chi \over 24}
\eea
where $\chi$ is the Euler characteristics of the 4-fold $T^8/G$.

Lifting to F-theory (or type IIB) the $H_{NSNS}$ and $H_{RR}$ fields
are
\bea
H_{NSNS}={1\over 2}(Ad{\bar z}^1dz^2d{\bar z}^3+Bdz^1d{\bar z}^2dz^3+
Cd{\bar z}^1dz^2
dz^3+Ddz^1d{\bar z}^2d{\bar z}^3)
\eea
\bea
H_{RR}={1\over 2i}(-Ad{\bar z}^1dz^2d{\bar z}^3+Bdz^1d{\bar z}^2dz^3+
Cd{\bar z}^1dz^2dz^3-Ddz^1d{\bar z}^2d{\bar z}^3)
\eea

Observe that

$\bullet$ All the field components have one of their legs along $z^3$.
Here
$z^3$ is the complex coordinate of $T^2/Z_2$.

$\bullet$ One can show that $H_{NSNS} \wedge *H_{RR}=0$. From the
equation
of motion
\bea
\nabla^2{\cal L}= -{3\over 2} H_{NSNS}\wedge *(H_{RR}-{\cal L}H_{NSNS})
\eea
would imply that
zero axion field, ${\cal L}=0$ is a consistent background. We
have also chosen
the dilaton $\Phi =0$.

$\bullet$ $\int H_{RR} \wedge H_{NSNS} = AB+CD = {\chi\over 24}$ which
is the required
for anomaly cancellation with $n=0$.

{}From the above equations its also easy to determine what the $B$
fields
would be {\it locally}:
\bea
B_{NSNS}={1\over 2}(A{\bar z}^1~dz^2d{\bar z}^3+Bz^1~d{\bar z}^2dz^3-
Cz^2~d{\bar z}^1
dz^3-D{\bar z}^2~dz^1d{\bar z}^3)
\eea
\bea
B_{RR}={1\over 2}(-A{\bar z}^1~dz^2d{\bar z}^3+Bz^1~d{\bar z}^2dz^3-
Cz^2~d{\bar z}^1
dz^3+D{\bar z}^2~dz^1d{\bar z}^3)
\eea

Therefore this is how we get the required background in type IIB.
One small subtlety is to see Lorentz invariance in full $3+1$
dimensions.
{}From the M-theory compactification it would seem like there is only
$2+1$ dimensional Lorentz invariance. However this is not the case as
can be
easily seen from the following arguments:

Compactifying from M-theory to IIA the
metric for the system
will have the following form:
\bea \label{metrforsyst}
\pmatrix{\Omega~g_{\mu\nu}&0\cr 0&\Omega'~ \eta_{\mu\nu}},
\eea
where $g_{\mu\nu}$ is the metric of a seven dimensional space  and
$\eta_{\mu\nu}$ is the $2+1$ dimensional Minkowski spacetime. Recall
that
the M-theory metric is related to IIA metric via some scaling implies
that
\bea \label{handphi}
\Omega = e^{{3\over 4}\phi} = \Omega'^{-1}
\eea
where $\phi$ is the dilaton. Now going from IIA to IIB undergoes an
inversion
of the T-dual direction which, from above equation, restores Lorentz
invariance in $3+1$ dimensions. On the remaining six dimensions we have
the
warp factor $\Omega$.
 The
six dimensional manifold in this case is $T^6/Z_2\times Z'_2$. However
the manifold we actually need is a deformed conifold. The case with a
trivial
fiberation over a conifold with fluxes can be worked out in some details
as
we saw in the previous section. Can that calculation be modified in
some
aspects so as to get the answer we require?

{\it Example 2}

We begin by  introducing  a circle action on the conifold and extend it
to the
resolved conifold and the deformed
conifold in a compatible manner \cite{dot,dot1}.

{\bf Conifold}:  Consider an action  $S_c$ on the conifold
$xy-uv=0$: \bea \label{sc}
S_c:~~(e^{i\theta}, x) \to x ,~~ (e^{i\theta}, y) \to y~~(e^{i\theta},
u) \to
e^{i\theta} u ,~~ (e^{i\theta}, v) \to
e^{-i\theta} v,~~ \eea The orbits of the action $S_c$ degenerates along
the
union of two intersecting complex
lines $y=u=v=0$ and $x=u=v=0$ on the conifold. Now, if we take a T-dual
along
the direction of the orbits of the action, there will be NS branes
along these
degeneracy loci as argued in \cite{bsv}. So we have two NS branes which
are
spaced along $x$ (i.e. $y=u=v=0$) and $y$  directions (i.e. $x=u=v=0$)
together with non-compact direction along the Minkowski space
 which will be denoted by $NS_x$ and $NS_y$.

{\bf Resolved Conifold}: This action can be lifted to the resolved
conifold.
 To do that, we
consider two copies of $\C^3$ with coordinates $Z, X, Y$ (resp. $Z',
X', Y'$)
for the first (resp. second) $\C^3$.
Then $\CO(-1) + \CO (-1)$ over $\P^1$ is obtained by gluing two copies
of $\C^3$
with the identification: \bea
\label{-1-1} Z' = \frac{1}{Z} ~, ~\quad X'= XZ ~, ~\quad Y' = YZ~. \eea
The $Z$
(resp. $Z'$) is a coordinate of
$\P^1$ in the first (resp. second) $\C^3$ and others are the
coordinates of
the fiber directions. The blown-down map from the resolved conifold
$\C^3 \cup
\C^3$ to the
conifold $\CC$ is given by \bea x=X=X'Z', ~~y=ZY=Y',~~u = ZX = X', ~~v=
Y =
Z'Y'. \eea {}From this map, one can see
that the following action $S_r$ on the resolved
conifold is an extension of the
action $S_c$ (\ref{sc}): \bea
S_r:~~(e^{i \theta }, Z) \to e^{i \theta} Z, ~~(e^{i\theta}, X) \to
X,~~
(e^{i\theta},
Y) \to e^{-i\theta}Y\nonumber \\
 (e^{i \theta }, Z') \to e^{-i \theta} Z', ~~(e^{i\theta}, X') \to
e^{i\theta}X',~~ (e^{i\theta}, Y') \to Y' \label{sr} \eea
The orbits degenerates along the union of two complex lines
$Z=Y=0$ in the first copy of $\C^3$ and $Z'=Y'=0$  in the
 second copy of $\C^3$.
Note that these two lines do not intersect and in fact they are
separated by the
size of
$\P^1$.
Now we take T-dual along the orbits of $S_r$ of type IIB theory. Again
there will be two NS branes
along the degeneracy loci of the action: one NS brane, denoted by
$NS_X$, spaced along $X$ direction (which is defined  by $Z=Y=0$ in the
first $\C^3$)
 and the other NS brane, denoted by $NS_{Y'}$  along
$Y'$ direction (which is defined by $Z'=X'=0$ in the second $\C^3$).
Here the length of the interval is the same as the
size of the rigid $\P^1$.  As the rigid $\P^1$ shrinks to zero, the
size of the
interval goes to zero
and $NS_X$ (resp. $NS_{Y'}$) approaches to $NS_x$ (resp. $NS_y$) of the
conifold.

{\bf Deformed Conifold}:
Finally we will provide a circle action of the deformed conifold and a
T dual
picture under this action. Consider the
following circle action $S_d$ \bea \label{sd} S_d:~~(e^{i\theta}, x) \to x
,~~
(e^{i\theta}, y) \to y,~~ (e^{i\theta},
u) \to e^{i\theta} u ,~~ (e^{i\theta}, v) \to e^{-i\theta} v ,~~ \eea
on the
deformed conifold \bea \label{conedeff} xy -uv =\mu
\eea Then $S_d$ is clearly the extension of $S_c$ (\ref{sc}) and the
orbits of
the action degenerate along a
complex curve  $u=v=0$ on the deformed conifold. If we take a T-dual of
the
deformed conifold along the orbits of
$S_k$, we obtain a NS brane along the curve $u=v=0$ with non-compact
direction
in the Minkowski space which is
given by \bea \label{dns} xy = \mu \eea in the x-y plane.
Topologically, the curve (\ref{dns}) is $\R^1 \times \S^1$.

{}From eq. (\ref{conedeff}) the metric is determined from Ricci flatness
and
Kahler potential ${\cal K}$
 as
\bea
ds^2 =  {\cal K'}~tr(dW^{+}dW)+{\cal K''} |tr(W^+dW)|^2
\eea
with $W$ satisfying $det~W = -{1\over 2} \mu^2$. The radial coordinate
in ${\bf C^4}$ is
\bea \rho^2 \equiv tr~(W^+W) \equiv \mu^2 cosh \tau \eea
The explicit metric for the deformed conifold is given as:
\bea \label{conidef}
ds^2= K \epsilon^{4\over 3}\Big( {sinh^3\tau \over 3(sinh~2\tau -
2\tau)}
(d\tau^2+ds_1^2) + {cosh~\tau\over 4} ds_2^2 + {1\over 4} ds_3^2 \Big)
\eea
where the quantities appearing in the above equation are defined in
\cite{candelossa,minatsim,ohtayokono}

As discussed in eq. (\ref{dns}) the T-dual of a deformed conifold is
given
in terms of two intersecting NS5 branes with a ``diamond'' structure at
the
center by using the terminology of \cite{kl}.
However a system of two intersecting NS5 branes are actually
delocalized in terms of supergravity solutions. Therefore, in this
limit,
we expect the following form of the deformed conifold metric:
\bea
ds^2 = A(r)^2dr^2+B(r)^2ds_1^2+C(r)ds_2^2+D(r)ds_3^2
\eea
As discussed in details in \cite{ohtayokono} the various coefficients
are
related as
\bea
B(r) = A(r)^{-1} = {1\over 4} C(r)^{-1} = {\delta \over 4} D(r)^{-1}
= f(r)^{-1}
\eea
where $|\delta| < 1$ is an integral constant and $f(r)$ is a linear
function
of $r$.

{}From the above discussion we conclude that an intersecting NS5 brane
configuration gives a conifold when we T-dualise along direction
$\theta
\equiv \psi \equiv x^6$. For the deformed conifold we will T-dualise
along direction $S_d$. This will be consistent with the way we
discussed the
deformed conifold.

Two intersecting NS5 branes intersecting at a point in the presence of
a
non constant $B_{NSNS}$ field along direction $\psi, \theta_2$ shows
a change in
the metric for the $\psi \psi$ and the $\theta_2 \theta_2$
components. If the value
of the dimensionless
$B_{NSNS}$ field at the origin is given by $b$ then
\bea
g_{\psi \psi} = {(1+b^2)f^2(r)\over 1+b^2f^3(r)} =
f(r) g_{\theta_2 \theta_2}
\eea
where $f(r)$ is a linear function of $r$. The $B$ field in this space
is
given by:
\bea
B = {bf^3(r)\over
1+b^2f^3(r)}(1+b^2)~(d\psi+cos\theta_1~d\phi_1+cos\theta_2
~d\phi_2) \wedge d\theta_2
\eea
This is basically the change in the background. To proceed further we
have to deform the intersection point of the two NS5 system. An
approximate
form of this has been worked out in \cite{ohtayokono}. In the absence
of
any fluxes the intersecting NS5 brane metric has an additional term
given by
\bea
2\beta f(r)~[{\cal A}~sin\alpha  + {\cal B}~cos\alpha]
\eea
where $\alpha$ is a constant, $\beta$ the size of the deformation and
\bea
{\cal A}=d\phi_1~d\theta_2~sin \theta_1+ d\phi_2~d\theta_1~sin
\theta_2\cr
{\cal B}=d\theta_1~d\theta_2 - d\phi_1~d\phi_2~sin \theta_1~sin
\theta_2
\eea
 This is
basically related to $ds_3^2$ term appearing in eq. (\ref{conidef}).
For
small values of $r$ the presence of $B$ field can be easily
incorporated
as a change in the $\psi \psi$ and $\theta_2 \theta_2$
 components of the new metric. To go to the
deformed conifold we have to T-dualise along the compact direction.
 We now perform the following transformation on the
coordinates $\theta_1,\psi$ as
\bea
\psi = \psi' cos~ \gamma, ~~~ \theta_1=\theta'~sec~\gamma + \psi'~
sin~\gamma
\eea
Using the known procedure we lift this configuration to M-theory along
direction $x^{10}$ and come
down back to type IIA via a different circle, say $x^7$. This will
create
a three form\footnote{Recall that because of this background
of four form field strength $G$ the susy
transformation will pick up an extra contribution of
$-{1\over 36} g_{\mu\nu} \Gamma_{\rho \sigma \lambda} G^{\nu \rho
\sigma
\lambda} \eta$ in $\delta_{\eta} \psi_{\mu}$.}
 in type IIA $C_{\psi' \theta_2,10}$ \cite{bergshoeff}.
Now making a T-duality along
$\psi'$ we essentially get a deformed conifold with a
\bea
B^{RR}_{\theta_2,10}=
{3\over 2} C_{\psi' \theta_2,10}, ~~~ B^{NSNS}_{\theta'\psi'} =
g_{\psi'\theta'}/g_{\psi'\psi'}
\eea
This is thus the required background in the presence of fluxes.

We remark that the delocalized solution is valid only for the case of
\cite{ks}, which is very close to the conifold and has a compact
direction
along which we can take a T-duality. For the case of \cite{vafa}, we cannot use
similar arguments because the T-duality is more complicated and we
cannot
take a T-duality in a clear space-time direction, although some
discussion in
this sense has been made in \cite{mt}.

\subsection{M-Theory with G-Fluxes: Exact Results}

Most of the discussions in the previous sections were motivated from
the
brane constructions and T-dualities. Though these techniques give us
the
back ground geometry, the fact that we are doing T-dualities introduces
{\it delocalization} in the picture. This is a major handicap. Question
is
can we improve upon this to get the exact background solution?  The
answer
turns out to be yes and we use the technique of M-theory on a fourfold
with G-fluxes. The fourfold we take is a non-trivial $T^2$ fiberation over the deformed conifold
$\cal B$, whose construction will be given in details later. As we discussed earlier, in this case we
expect the metric of the fourfold to be warped.

The warped metric can be written again as:
\bea
ds^2=e^{-\phi(y)}\eta_{\mu\nu}dx^{\mu}dx^{\nu}+e^{{1\over
2}\phi(y)}g_{a
\bar b}dy^ady^{\bar b}
\eea
where $g$ is the metric for the unwarped internal fourfold which is
parametrised by complex coordinates $y^a,y^b$ (or real coordinates
$y^m$)
and $\eta$ is the three
dimensional space parametrised by $\mu, \nu$. The eleven dimensional
spinor
$\kappa_0$ decomposes as
\bea  \kappa_0 = \epsilon \otimes \zeta \eea
and the gamma matrices decompose as
\bea \Gamma_{\mu}= \gamma_{\mu} \otimes \gamma_9, ~~~~ \Gamma_m = 1
\otimes
\gamma_m \eea
where $\gamma_9$ is the eight dimensional chirality operator. Here
$\epsilon$ is a three dimensional anticommuting spinor and $\zeta$ is a
commuting eight dimensional Majorana-Weyl spinor.

{}From the susy variation
of the gravitino $\delta \psi_{\mu}=0$ we expect to find a spinor
satisfying
\bea \label{onepsil}
\nabla_{\mu}\epsilon = 0
\eea
The existence of a covariantly constant spinor puts various constraints
on the background.

\noindent{\bf 1. The Five-form equation:}
Eq. (\ref{onepsil}) imposes the following
constraints on the background four form
$G = dC$:
\bea \label{consfour}
e^{3\phi/2} (\gamma_{\mu} \otimes \gamma^m) \epsilon^{\mu \nu \rho}
G_{\mu \nu \rho m} \kappa - {3\over 2} \del_n \phi (\gamma_{\mu}
\otimes
\gamma_9 \gamma^n) \kappa =0 \cr
\Rightarrow G_{\mu \nu \rho m} = \epsilon_{\mu \nu \rho}\del_m
e^{-{3\phi
\over 2}} \eea
Observe that this G-flux is generated entirely from the warp factor.
When we shrink the size of the fibered torus to zero then this four
form of
M-theory goes to five form ${\cal F}$ of F-theory (or IIB). The two
three form fluxes are
respectively the NSNS and RR fluxes which come from the {\it
background}
fluxes switched on as:
\bea
\omega- *_{\cal B}~\omega, ~~~~~ \omega\tau - *_{\cal B}~ \omega {\bar \tau}
\eea
This can be related to the five form as
\bea \label{lineq} {\cal F} = {3\over 4}\epsilon^{ij}~*~(B^{(i)}\del
B^{(j)})
\eea
with $i,j=1,2$ being the two $B$ fields. The Hodge $*$ here is wrt
the ten dimensional metric.
The above equation can be argued
from the self-duality condition of the modified five-form of type IIB theory.
Observe that this is precisely
one of
the linear equation relating the derivative of the warp factor to the
background $B$-fields as discussed in \cite{ks}.

\noindent{\bf 2. The Donaldson-Uhlenbeck-Yau equation:}
Our discussion regarding eq. (\ref{onepsil}) is however not complete.
There is
yet another condition \cite{bbsisters,gukov,drs,guk} on the background
four-form. This is
the Donaldson-Uhlenbeck-Yau equation for the fourfold which puts a
constraint
 on the G-flux as
\bea \label{consgflux}
G_{a {\bar b} c {\bar d}}~ g^{c{\bar d}} = 0
\eea
where $g$ is the metric of the fourfold.
If $z$ parametrizes the fiber torus and $\cal B$ the deformed conifold
base
whose complex coordinates are $a,b$ then the above equation gives us
the following set of equations:
\bea G_{a {\bar b} z {\bar d}}~ g^{z{\bar d}} = 0,
~~~~~ G_{z {\bar a} b {\bar d}}~ g^{b{\bar d}} = 0 \eea
which is nothing but the self duality relations of the G-fluxes. This
further implies
\bea \label{lineq2} H_{NSNS} = *_{\cal B}~ H_{RR} \eea
This equation relates the NSNS and RR three form field strengths
linearly
giving us the other equations of \cite{ks}.

{\it F-theory and  Orientifold Limits}

In the previous sections we avoided a subtlety regarding the global
structure of the system in IIB and we now turn to clarify that issue.
We will construct an elliptically fibered Calabi-Yau fourfold over a
compactification of the deformed conifold.
 For the fourfold
which is a non-trivial $T^2$ fiberation over a base ${\cal B}$ $-$ with the
$T^2$ degenerating at some points on the base $-$ this situation is similar
to the case discussed in \cite{drs}. Looking from F-theory point of view,
as discussed earlier in some details, an
F-theory compactification on an elliptically fibered Calabi-Yau
fourfold is equivalent
to type IIB string theory on the base of the fiberation, where
the type IIB coupling is identified
with the modular parameter of the elliptic curve.

We begin with the deformed conifold defined
in $\C^3$ by
\bea
\label{defcon}
 x_0^2 + x_1^2 + x_2^2 + x_3^2  = \mu.
\eea
 For non-zero $\mu$, this is smooth and
symplectically isomorphic to
the cotangent bundle $T^*\S^3$ over the three sphere $\S^3$. As $\mu \to 0$, the compact cycle
$\S^3$ will vanish and the conifold singularity will develop.
By adding $\P^1 \times \P^1$ to the boundary of (\ref{defcon}), we can
compactify to a projective
variety $\CB_\mu$ in $\P^4$ defined by a quadratic equation:
\bea
\label{quadric}
\CB_\mu:~~ z_0^2 + z_1^2 + z_2^2 + z_3^2  - \mu z_4^2 =0.
\eea
The quadric threefold $\CB_\mu$ does not develop any new singularities at
infinity and thus
smooth for $\mu \neq 0$ and has a conifold singularity at $(0,0,0,0,1) \in \P^4$ when
$\mu=0$. Moreover, the anti-canonical bundle \bea -\CK_{\CB_\mu} :=
\wedge^3 {\bf T}\CB_\mu \eea will be the restriction of $\CO(3)$ of $\P^4$
to $\CB_\mu$ by the adjunction formula,  and hence it is very ample. From
Kodaira vanishing theorem, one can also show that $H^i(\CB_\mu, \CO) =0$ for $i>0$.
We now define a fourfold
$Y_\mu$ as a subvariety in  the projective bundle
$\P ({\cal O} \oplus {\cal L}^2  \oplus {\cal L}^3)$ where $\CL := \CK_{\CB_\mu}^{-1}$,  given by the
Weierstrass equation
\bea
y^2z = x^3 + f z^2x + g z^3,
\eea
where $z, x,y, f, g$ are the sections of ${\cal O}, {\cal L}^2, {\cal L}^3,
{\cal L}^4, {\cal L}^6$ respectively.
Since the anti-canonical bundle
${\cal L} = {{\cal O}(3)}|_{\CB_\mu}$ is very ample,
we may choose $f$ and $g$ so that the fourfold $Y_\mu$ is smooth.  By the
projection formula, one can see that $Y_\mu$ is Calabi-Yau since $H^i( \CB_\mu,\CO) =0, i>0$.
By construction, the natural projection \bea \P ({\cal O} \oplus {\cal L}^2  \oplus {\cal L}^3) \to     \CB_\mu \eea induces a
fiberation $Y_\mu \to \CB_\mu$ whose fibers are elliptic curves.
F-theory on the Calabi-Yau fourfold $Y_\mu$ is by definition type IIB
theory
compactified on the base $\CB_\mu$ with background axion-dilaton field
$\lambda$ whose $j$-invarinat is given by:
\bea
j (\lambda) = \frac{ 4 \cdot (24 f)^3}{4 f^3 + 27 g^2}.
\eea
and various $(p,q)$ seven branes appearing at the loci where the elliptic
fiberation degenerates, i.e. where the discriminant
\bea
\label{disc}
\Delta = 4 f^3 + 27 g^2
\eea
vanishes.
Using Riemann-Roch for
$\CB_\mu$ and integrating over the elliptic fibers, one can evaluate
the Euler-Characteristic $\chi$ of $Y_\mu$
\bea
-Q_3^{D7} = \frac {\chi}{24} = 12 + 15 \int_{\CB_\mu} c_1(\CB_\mu)^3 =  822.
\eea
as in \cite{sethi}.
{}From the above consideration we  expect that the
warped
F-theory compactification on $Y_\mu$ can be related to
M-theory on the Calabi-Yau
fourfold
with G-fluxes as discussed in details in
\cite{bbsisters,drs,gubser,kachru}. In IIB we therefore get a
set of D7 branes and $O7$ planes along with a deformed conifold
background and fluxes. However the issue of fluxes is more
subtle now because of the presence of branes and orientifold planes.
There are two interesting cases we have to consider from M-theory point of
view:

\noindent 1. The G-fluxes are localised at points where the $T^2$ fiber
degenerates.

\noindent 2. The G-fluxes are spread over the fourfolds but with a
normalisable $\int G\wedge G$.

{}From the first case we will have a decomposition
\bea G = \sum_{i=1}^k F_i \wedge [\Omega_i]\eea
where we take the fiber degenerating at $k$ points and $[\Omega_i]$ are
normalisable two forms localized at the singularities. Since the
branes are located at those points, we see that the background G-fluxes
have actually appeared as gauge fluxes on the branes\footnote{Recall that
the orientifold planes are $(p,q)$ 7 branes.}. The warp factor equation
will also be different now since both branes and planes are sources of
$tr (R \wedge R)$\cite{Greendjat}. The warp factor is given as:
\bea
\Delta e^{3\phi/2} = *_{\cal B}\sum_{i=1}^k [F_i\wedge F_i + tr(R\wedge R)]
\delta^2(l-l_i) \eea
where $l_i$ are the positions of the branes and planes.

For our purpose however we require the second condition wherein we
decompose\footnote{The two $B$-fields survive the orientifold projection.}
 the background G-flux as
\bea
{G\over 2\pi} = dz \wedge \omega - d {\bar z} \wedge *_{\cal B}~ \omega
\eea
In the limit when the size of the three cycle of the base goes to zero
this
fourfold will be related to the one discussed in \cite{kachru}.
Therefore
following the discussions of \cite{senorient,kachru}, near the conifold
point $\mu \to 0$, we can study the IIB
background as though we are removed far away
from
the $O7$ planes and D7 branes. In this limit the calculations of the
previous
sections will give us exact results in the local neighbourhood of the
singularity \cite{kachru}.

\subsection{The Transition From G-fluxes}

As discussed in detail in section (3.2) the type IIB metric is given as
\bea \label{metrforsysttwo}
\pmatrix{\Omega~g_{\alpha \beta}&0\cr 0&\Omega'~ \eta_{\mu\nu}},
\eea
where $\mu,\nu$ runs over the $3+1$ dimensional spacetime and
$g_{\alpha \beta}$ is the metric of the deformed conifold. $\Omega$ is
the
warp factor and we can use it to write the ten dimensional IIB metric
explicitly as
\bea \label{iibmetric}
ds^2 &=& e^{-{3\phi \over 4}} dx_{\mu}dx^{\mu} +e^{3\phi \over 4}
g_{\alpha \beta} dx^{\alpha} dx^{\beta}\cr
&=& H^{-{1\over 2}}dx_{\mu}dx^{\mu}+H^{1\over 2}
g_{\alpha \beta} dx^{\alpha} dx^{\beta}
\eea
where we have used eq. (\ref{handphi}):
\bea
\Omega = e^{3\phi \over 4} = \Omega'^{-1} \equiv H^{1\over 2}
\eea
As discussed in \cite{ks} the above form of the metric is in the same
category as a D-brane metric. We thus see that eq. (\ref{iibmetric})
can
be derived from M-theory. From the above consideration is it possible
to
see the geometric transition to the wrapped brane picture?

For this first let us consider a small
resolution of the compactified conifold $\tilde{\CB_0} \to \CB_0$.
We can pull back the elliptic fiberation $Y_0 \to \CB_0$ to
$\tilde{\CB_0}$ which will be denoted by
$\tilde{Y_0}$.
Hence the conifold transition can be lifted to the Calabi-Yau fourfold
transition:

\vskip.5in

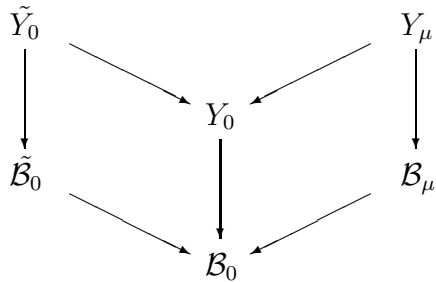
\setlength{\unitlength}{0.4mm}
\begin{picture}(400, 80)(-62,0)
\put(60, 83){$\tilde{Y_0}$}
\put(60, 33){$\tilde{\CB_0}$}
\put(125, 53){$Y_0$}
\put(125, 3){$\CB_0$}
\put(190, 83){$Y_\mu$}
\put(190, 33) {$\CB_\mu$}
\put(80, 30){\vector(2,-1){40}}
\put(80, 80){\vector(2,-1){40}}
\put(180, 30){\vector(-2,-1){40}}
\put(180, 80){\vector(-2,-1){40}}
\put(65, 78){\vector(0,-1){33}}
\put(130, 48){\vector(0,-1){33}}
\put(195, 78){\vector(0,-1){33}}
\end{picture}
\\
\newline
\noindent
In  the fourfold transition, the real five dimensional cycle in $Y_\mu$
will
shrink to zero as $\mu$ approaches to zero and the four dimensional
cycle will blow up in $\tilde{Y_0}$.
How does the background warped metric transform under this?

\noindent{\bf Conifold transition of the base}:
The base $-$ deformed conifold $-$ metric has a form given in
eq. (\ref{conidef}) as
\bea
ds^2 = A(\tau) (d\tau^2+ds_1^2) + B(\tau) ds_2^2 + C(\tau) ds_3^2
\eea
where the quantities $A,B,C$ are defined earlier. For this case there
are
two different limits:
\bea
(1)& \tau \to 0,& \mu = \mbox{ fixed} \\
(2)& \tau \to \infty,&  \mu \to 0, ~~~~ \mu~cosh~\tau=
\mbox{ fixed}
\eea
The first case tells us that the deformed conifold reduces to a $S^3$
and in
the second case the metric reduces to the conifold with the radius
parameter given by $(\mu~cosh~\tau)^{1\over 3}$. In this limit
the
equation of the warp factor
\bea
\nabla_{\mu}{\cal F}^{\mu \mu_1 \mu_2 \mu_3 \mu_4} =
{3\over 5!4}\epsilon^{ij}\epsilon^{\mu_1 \mu_2 .... \mu_{10}}
H^{(i)}_{\mu_5 \mu_6 \mu_7} H^{(j)}_{\mu_8 \mu_9 \mu_{10}}
\eea
and eq. (\ref{iibmetric}) will give us the UV metric of \cite{kn}\footnote{We
are again assuming that the branes and planes don't alter the results in
any substantial way.}.
Far
in the IR (for the first case) when we make $S^3 \to 0$ there would be
{\it source} for the $H_{RR}\equiv H'$ signifying the presence of a wrapped D5.
A way to argue this is from the background equation of motion for the
RR fields:
\bea \label{hrr}
\nabla^{\mu}H'_{\mu\nu\rho} = (*{\cal F})_{\nu\rho\sigma\lambda\kappa}
H'^{\sigma\lambda\kappa} {\Big |}_{background} \eea
where we have assumed a constant axion\footnote{We are taking a
limit in which $B_{NSNS} \to \epsilon$
and $B_{RR} \to \epsilon^{-\beta}$. Therefore from the above equation we have
$\nabla H' \to \epsilon^{1-2\beta}$. The fact that $\beta > 1$ can be argued
from the finiteness of the $H'$ flux. Observe that in the far IR
$B_{NSNS}\to 0$ so there will be no contribution to the four form charge
$\int H_{NSNS}\wedge H_{RR}$.}.
An alternative way to see that that there is a wrapped brane is to go to
the T-dual picture of the deformed conifold. At the intersection, in the limit
of the vanishing size of the diamond, there is a strong flux which creates
a brane by a mechanism similar to the Hanany-Witten effect\cite{dot}.

The above analysis is therefore a way to see the transition for the
Klebanov-Strassler case. Thus at the conifold point we have a
wrapped
D5 and a $B_{NSNS}$ field. This $B_{NSNS}$
field
will appear in the $3+1$ dimensional gauge theory as a coupling
constant.
Vafa's case would be to trade the varying $B_{NSNS}$ field with the
size
$\cal Z$ of a
blown up $P^1$. For our purpose we can define a coupling constant
\bea
\tilde{\tau} = \int B_{NSNS} + i {\cal Z} \eea whose $Re ~\tilde{\tau}$ and
$Im~\tilde{\tau}$ determine essentially the two model. In fact this is
related to the brane construction we had discussed in the earlier
sections.
For the Klebanov-Strassler case we had two intersecting NS5 brane
separated
along $x^6$ and a D4 brane stretched between them. For Vafa's case the
system is separated along $x^7$. The most general model would be to
stretch
the two NS5 on a complex plane $z \equiv x^6 + i x^7$.

We discussed in this section the solution of \cite{ks} obtained from
F-theory
compactified on a non-trivial fiberation over the deformed conifold,
where the
warp factor is non-constant. We could ask now what happens if we consider
Vafa's model \cite{vafa}. From the above discussions we see that as long as
we take a {\it non-constant} size of the $P^1$ in the resolved side we in fact
generate an identical back ground after the conifold transition. Therefore
we would expect the warp factor to be same as the Klebanov-Strassler
case. However if the size of the $P^1$ is a constant then ${\cal F}=0$
resulting in a constant warp factor. In general the transition is to a
non-constant warp factor determined mainly by the NSNS field.

\section{\bf The Nonsupersymmetric Background}

Existence of a supersymmetric background is equivalent to saying that we
have a primitive $(2,2)$ form $f_{(2,2)}$\cite{bbsisters,drs}. The
generalised primitivity condition is defined on a $(p,q)$ form as
\bea J \wedge f_{(p,q)} = 0 \eea
When supersymmetry is
completely broken then we can have (2,2) form which is not primitive
and is
given in terms of a closed (0,0) form $f_{(0,0)}$ and the kahler form
$J$
as $J\wedge Jf_{(0,0)}$ \cite{bbsis}. Of course this form is in
addition to
the primitive (2,2) form and the (0,4) and (4,0) form. The complete
four
form background that can be switched on for the non supersymmetric case
is
\bea G = f_{(4,0)}+f_{(0,4)} + f_{(2,2)} + J \wedge J f_{(0,0)}\eea It was
shown in \cite{bbsis} that in these models the three dimensional
cosmological constant vanishes.

The non primitive $(2,2)$ form also receives contributions from $(1,1)$
form as $J \wedge f_{(1,1)}$\cite{gukov,guk,haack}. Using the
tadpole cancellation
condition \bea \int_{Y} G\wedge G = {\chi \over 24} \eea
it was shown in \cite{gukov,guk,haack} that we generate
a potential in three dimensions
as \bea V = \int_{Y}~ |G|^2 - {\chi\over 6} \eea
where a $(3,1)$ and $(1,3)$ background is also switched on a Calabi-Yau
fourfold $Y$ which has $T^2$ fiberation over a Fano threefold $\cal B$.

The above discussions therefore indicates that we can extend our results to
the non susy case also. It remains however to see what happens after the
transition. As pointed out to us by S. Kachru \cite{shamit} this susy
breaking is more of a global effect and therefore it mayn't be possible to
argue susy breaking in the dual field theory. However the examples
considered above are for compact fourfolds. For non-compact cases the
situation will be different.
More details on this will be reported elsewhere.

\section{\bf Discussions}

\subsection{Is there a Type I dual also?}

In this paper we have given a M-theory compactification which, in some
limits, reproduce the Klebanov-Strassler or Vafa's model. However there are
many interesting directions still remain to be explored. Let us go back again
to our  construction of the fourfold. The deformed conifold has been
compactified to a smooth projective threefold in $\P^4$:
\bea
{\cal B}_\mu:~~ z_0^2 + z_1^2 + z_2^2 + z_3^2 - \mu z_4^2 =0.
\eea
Now consider a family ${\cal F}$ of $\P^2$'s in $\P^4$ which contain a fixed generic projective line
$l_0$
at the infinity (i.e. $z_4 =0$). So here we assume that the line $l_0$ will
intersect $\CB_\mu$ at two distinct points.
The family $\CF$ is two dimensional and
parametrized by $\P^2$. So after blowing up the line $l_0$ in $\P^4$, we
obtain a $\P^2$ fibered space $\tilde {\P^4}$ over the parameter space $\P^2$.
Let $\tilde{\CB_\mu}$ be the proper transform of $\CB_\mu$ under this
blow-up. Then we have a family of quadric curves over $\P^2$:
\bea
\pi :\tilde{\CB_\mu} \to \P^2.
\eea
The quadric curves are
 obtained by intersecting $\CB_\mu$ with
$\P^2$ and isomorphic to $\P^1$.
Now we choose a smooth generic quadric threefold $Q$ in $\P^4$
which will
intersect with $\CB_\mu$ smoothly.
As  a special case of F-theory constructed in section 4.3, we may assume
that the discriminant (\ref{disc}) is of the form
\bea
\Delta = 4 f^3 + 27 g^2 = h^{18}
\eea
with $h \in \CO(2)$.
Let $Q$ be the quadric threefold in $\P^4$ defined by $h$.
We take a double covering $W$ over $\P^4$ which is ramified over the
smooth divisor $Q$. Then the fiber product $\tilde{\P^4} \times_{\P^4}
W$ will be a double cover over $\tilde{\P^4}$ ramified over the proper
transform $\tilde{Q}$ of $Q$. We restrict the double covering
\bea
f: \tilde{\P^4} \times_{\P^4} W\to \tilde{\P^4}
\eea
to the inverse image $\CE$ of $\tilde {\CB_\mu}$ under $f$. Therefore
 $\CE$ will be
a double cover of $\tilde{\CB_\mu}$ ramified over
$\tilde{Q}\cap \tilde{\CB_\mu}$. Hence we have the following situation:
\\
\setlength{\unitlength}{0.4mm}
\begin{picture}(400, 80)(-70,0)
\put(70, 53){$\CE$}
\put(120, 53){$\tilde{\CB_\mu}$}
\put(120, 3){$\P^2$}
\put(95,61){$f|_\CE$}
\put(130, 30){$\pi$}
\put(85, 56){\vector(1,0){30}}
\put(125, 48){\vector(0,-1){35}}
\end{picture}
\\
\newline
\noindent
The general fibers of the composite map $f|_\CE \circ \pi$ will be
elliptic curves since they are double covering of $\P^1$ ramified over 4
points. Thus the map
\bea
f|_\CE \circ \pi : \CE \to \P^2
\eea
will be a $T^2$ fiberation and there is an involution $\CI_2$ such that
$\CE / \CI_2 = \tilde{\CB_\mu}$.
Here the  72 branes are grouped into four sets of 18
coincident branes situated on the ramification divisor
$\tilde{Q}~ \cap~ \tilde{\CB_\mu}$ and are located
at the fixed points of
$\CI_2$.
So as in \cite{sen97}, we have type IIB compactification on
$\tilde{\CB_\mu}$
such that we go once around each fixed point of $\CI_2$ the theory comes
back to itself transformed by the symmetry $ (-1)^{F_L}.~\Omega~$
where $(-1)^{F_L}$ changes the sign of all the Ramond sector states on the
left moving fermions
and $\Omega$ denotes the orientation reversal transformation. Therefore
the theory can be identified to type IIB on $\CE$, moded out by the
$\Z_2$ transformation
\bea   (-1)^{F_L}.~\Omega~.~{\CI_2}. \eea
By making  T-dualities along both the circles of the $T^2$, we can map the
$\Z_2$ transformation $(-1)^{F_L}.~\Omega~.~{\CI_2}$ to $\Omega$.
Since modding out the type IIB theory by $\Omega$ produces type I theory,
we obtain type I theory on $\CE$~\cite{sen97}. By pulling back $H$ fields
constructed before on $\tilde{\CB_\mu}$
to $\CE$,
we expect the $H_{NSNS}$ fields to dissolve in the
metric in type I background\footnote{Those which don't are actually projected out by the
orientifold projection.} and the metric will become non-K\"ahler i.e.
the metric will have torsion\footnote{A similar case has been noticed earlier
in \cite{drs}.}.
The RR field
will appear as a three form field in
type I. Notice that we have changed a compactification of the deformed
conifold  from $\CB_\mu$ to $\tilde{\CB_\mu}$ which is still Fano so that
the  discussions in the previous sections go through.
As discussed in \cite{smitandy86,drs} the three
forms $H$ are related to
the Type I metric as
\bea \label{100}
H = {i\over 2}({\bar \del}-\del)~J \eea
Here $J$ is the $(1,1)$ form associated with the metric which becomes
non-K\"ahler due to $H_{NSNS}$ after T-dualities.
In terms of components the above equation can be recast as
\bea \label{101} H_{{\bar a}{\bar b}c}= -g_{c[{\bar a},{\bar b}]},~~~
H_{ab{\bar c}}= -g_{{\bar c}[a, b]}\eea
where $[~~]$ denotes antisymmetrisation.
The equation (\ref{101}) can be shown to
reproduce the linear equation  written earlier as eq. (\ref{lineq})
when we suppress the $X_8$ and the membrane terms.
 In order to see this, one only
needs to consider the components of $H_{NSNS}$  and $H_{RR}$ fields
which have one leg along the $T^2$
direction as the other components are projected out.
Then we can show that the self-duality of G-flux (\ref{lineq2}) is
equivalent to
(\ref{100}).  This is not surprising considering the fact that both
conditions are derived in order to have supersymmetry.
The warp factor for $\CE$  will descend to $\CB_\mu$ because $H$ fields have been lifted from
$\CB_\mu$.
Therefore this serves as
another alternative way to see the background  equation of motions.
However the above method takes into account the regions close to the D7-O7
systems.
But for the Klebanov-Strassler case we have restricted ourselves to the regions
far away from the D7-O7 system.
It would therefore be interesting to see the
range of validity of the above technique.

\subsection{Fate of the open strings}

Another interesting issue here is the fate of the open strings. In the final
picture we have a complete closed string background without any open strings.
In the dual brane picture the open strings on the D4 branes become 2-branes
when the system is taken to strong coupling. They then combine with the
two branes which determine the dynamics of the MQCD five brane. This system
eventually becomes two intersecting M5 with a diamond structure which is
a closed string background in IIB.

An interesting question is to see whether we can argue {\it directly} the
existence of open strings in the background described by eq. (\ref{hrr}).
To do this we have to determine the zero mode fluctuation of the background
three forms. Let us decompose the three form as
\bea H_{RR} = H_{RR} {\Big |}_{background} + h \eea
where $h$ is the fluctuation. As we argued earlier in eq. (\ref{hrr}) the
limit when $S^3 \to 0$ we have a source of a D5 brane. How does the
fluctuation manifest itself now?

To see this we need to recall the conifold equation written suggestively as
\bea
x_1^2+x_2^2+x_3^2 = - x_4^2 \eea
In this form it describes a ${\bf Z_2}$ ALE space fibered over a $x_4$
plane. We can use this information now to our advantage. Recall that a
${\bf Z_2}$ ALE space supports a normalisable harmonic two form $l_2$.
Therefore the small fluctuations of the background three form field can be
decomposed as
\bea h = A \otimes l_2 \eea
where $A$ is a one-form restricted on the plane $x_4$ and spacetime. Therefore
this fluctuations appear as $U(1)$ gauge fields on the D5 world-volume. This
may be one way to understand the transition from closed string backgrounds
to open string backgrounds.

Before we end, observe that the closed string background is a
conjectured
dual \cite{ks}  of the ${{\cal N}=1}$ pure glue
theory in the following limits:
\bea g_s \to \epsilon, ~~\int_{S^2\to 0} B_{NSNS}\to \epsilon \eea
Observe that the above limit is {\it opposite} of the limit discussed in
eq. (\ref{kslimit}) for the validity of the sugra backgrounds. This implies
that we really don't have a rigorous proof of the dual of the pure glue
theory from supergravity point of view. However as we saw earlier the
brane construction method are at {\it finite} $g_sM$. And
therefore they provide a strong
argument in the favor of this duality.

\section{Acknowledgement}

We would like to thank Donu Arapura, Antonella Grassi, Aki Hashimoto,
Igor Klebanov,
Juan Maldacena, Prabhakar Rao  and Nathan Seiberg
for many helpful discussions
during the course of this work. We would also like to thank Shamit Kachru
and Dieter L\"ust for useful comments on the draft.
KO thanks IPAM for hospitality during the
last
stage of this work.
The research of KD is supported in part by David and Lucile Packard
Foundation Fellowship $2000-13856$.
The research of KO is supported in part by UM Research Board.
The research of JP is supported in part  by POSTECH BSRI fund 1RB011601.
The research of RT is supported by DFG.

\end{document}